\documentclass[aps,prb,floatfix,twocolumn,superscriptaddress]{revtex4-2}

\usepackage{mathrsfs}
\usepackage{graphicx}
\usepackage[english]{babel}
\usepackage{amsmath}
\usepackage{amssymb}
\usepackage{xcolor}
\usepackage{cancel}
\usepackage[normalem]{ulem}

\usepackage{relsize}
\usepackage{CJK}
\usepackage{dsfont}
\usepackage{bm}

\newcommand{\me}{\mathrm{e}}
\newcommand{\mi}{\mathrm{i}}

\newcommand{\dif}{\mathrm{d}}

\allowdisplaybreaks

\begin{document}
\title{Interferometric Geometric Phases of $\mathcal{PT}$-symmetric Quantum Mechanics}

\author{Xin Wang}
\affiliation{School of Physics, Southeast University, Jiulonghu Campus, Nanjing 211189, China}

\author{Zheng Zhou}
\affiliation{School of Physics, Southeast University, Jiulonghu Campus, Nanjing 211189, China}

\author{Jia-Chen Tang}
\affiliation{School of Physics, Southeast University, Jiulonghu Campus, Nanjing 211189, China}

\author{Xu-Yang Hou}
\email{houxuyangwow@seu.edu.cn}
\affiliation{School of Physics, Southeast University, Jiulonghu Campus, Nanjing 211189, China}

\author{Hao Guo}
\email{guohao.ph@seu.edu.cn}
\affiliation{School of Physics, Southeast University, Jiulonghu Campus, Nanjing 211189, China}
\author{Chih-Chun Chien}
\affiliation{Department of physics, University of California, Merced, CA 95343, USA}
\begin{abstract}
We present a generalization of the geometric phase to pure and thermal states in  $\mathcal{PT}$-symmetric quantum mechanics (PTQM) based on the approach of the interferometric geometric phase (IGP). The formalism first introduces the parallel-transport conditions of quantum states and reveals two  geometric phases, $\theta^1$ and $\theta^2$, for pure states in PTQM according to the states under parallel-transport. Due to the non-Hermitian Hamiltonian in PTQM, $\theta^1$ is complex and $\theta^2$ is its real part. The imaginary part of $\theta^1$ plays an important role when we generalize the IGP to thermal states in PTQM. The generalized IGP modifies the thermal distribution of a thermal state, thereby introducing effective temperatures. At certain critical points, the generalized IGP exhibits discrete jumps at finite temperatures, signaling a geometric phase transition. We demonstrate the finite-temperature geometric phase transition in PTQM by a two-level system and visualize its results.
\end{abstract}

\maketitle

\maketitle

\section{Introduction}
The introduction of non-Hermitian quantum mechanics (NHQM) \cite{FESHBACH1958357,Moiseyev_book,El_Ganainy_2018} has uncovered many fascinating phenomena, including the Anderson localization~\cite{PhysRevLett.77.570}, gapless quantum phase transitions \cite{Matsumoto}, unconventional behavior of quantum emitters \cite{GZ22,Roccati_2022}, tachyonic dynamics \cite{Liegeois2022,PhysRevLett.98.253005}, and distinctive topological properties \cite{PhysRevX.8.031079,PhysRevLett.121.026808,Roccati2023}. A major branch of NHQM includes systems with non-Hermitian Hamiltonians obeying parity-time reversal ($\mathcal{PT}$) symmetry, which can possess real-valued eigenvalues, making them a relevant extension of conventional quantum mechanics.
Therefore, $\mathcal{PT}$-symmetric quantum mechanics (PTQM) has attracted considerable research attention in many aspects \cite{PhysRevLett.80.5243,PhysRevB.82.052404,PhysRevLett.110.083604,Korff_2007,Korff_2008,RevModPhys.88.035002,PhysRevA.106.032206,Bender24} and has been experimentally realized across different fields, including acoustics, optics, electronics and quantum systems \cite{Cham_2015,Feng_2017}. 
It has also catalyzed extensive investigations into physical and topological characteristics of non-Hermitian systems \cite{Ashida_2017,PhysRevLett.119.190401,Weimann_2016,PhysRevB.95.174506,PhysRevB.98.085116,PhysRevB.98.085126,PhysRevLett.120.146402,PhysRevLett.121.136802,Fritzsche2024}.

Geometric phases in quantum systems, including the Berry phase \cite{Berry84} and the Aharonov-Anandan phase \cite{PhysRevLett.58.1593}, have advanced our understanding of the geometric structures behind interesting physical systems and shown significant influence across various fields. For instance, the Berry phase is fundamental in the study of topological matter since it connects geometric objects from the underlying mathematical structure to measurable physical quantities~\cite{TKNN,Haldane,KaneRMP,ZhangSCRMP,MooreN,KaneMele,KaneMele2,BernevigPRL,MoorePRB,FuLPRL,Bernevigbook,ChiuRMP}. Recently, the notion of geometric phase has been generalized to non-Hermitian quantum systems \cite{JPAGW13}, which is further applied to the construction of the quantum geometric tensor for non-Hermitian systems \cite{Zhang2019}. On the other hand, the geometric phase has also been generalized to mixed quantum states via different approaches~\cite{GPMQS1,GPMQS2,GPMQS3,GPMQS4,GPMQS5,GPMQS6,Andersson13,DiehlPRX17,WangSR19}. In this work, we will generalize the one proposed by Sj\"oqvist et al \cite{PhysRevLett.85.2845} based on an extension of the optical process in the Mach-Zehnder interferometer, which is referred to as the interferometric geometric phase (IGP). Numerous studies \cite{PhysRevA.67.020101,PhysRevLett.90.160402,Faria_2003,PhysRevLett.93.080405,Kwek_2006,Andersson_2016} have been dedicated to this field, and the IGP has been observed by various techniques, including the nuclear magnetic resonance \cite{PhysRevLett.91.100403,Ghosh_2006}, polarized neutrons \cite{PhysRevLett.101.150404}, and the Mach-Zehnder interferometer \cite{PhysRevLett.94.050401}.
A different approach was introduced by Uhlmann \cite{Uhlmann86,Uhlmann89,Uhlmann1992} soon after the discovery of the Berry phase and the phase is usually called the Uhlmann phase. This approach incorporates a full mathematical structure based on fiber bundles and has gained attention due to its relevance to condensed matter and quantum information~\cite{Viyuela14,ViyuelaPRL14-2,ourPRB20,OurPRB20b,Galindo21,Zhang21,OurPRA21,Hou2023}.

We aim at generalizing the concept of IGP to thermal states in PTQM. In the beginning, we establish a formalism for pure-state geometric phase in $\mathcal{PT}$-symmetric systems by using the conventional derivation and  then introducing the parallel-transport conditions of quantum states. In contrast to conventional QM, a PTQM system is shown to allow two distinct geometric phases, called $\theta^{1}$ and $\theta^2$, which are differentiated by the states undergoing parallel-transport. On the one hand, $\theta^2$ exactly coincides with the known result in Ref.~\cite{JPAGW13} and is the real part of $\theta^1$. On the other hand, $\theta^1$ is a complex-valued phase with its imaginary part  adjusting the amplitude of the wavefunction due to the lack of Hermiticity. Since $\theta^1$ will be shown to be associated with the non-Hermitian Hamiltonian, the generalization to thermal states in PTQM will be based on it.

Following the construction of the IGP of thermal states in conventional QM and the derivation of the geometric phase $\theta^1$, we develop a framework of the IGP of thermal states in PTQM. In general, the IGP is the argument of the thermal-weighted sum of the geometric phase factor for each individual energy level. The imaginary part of the generalized IGP will be shown to alter the relative thermal weights, which introduces effective temperatures to the thermal states. Consequently, there may be quantized jumps of the IGP at certain temperatures and  system parameters.
This phenomenon signifies a geometric phase transition at finite temperature in PTQM. To illustrate our findings and visualize the results, we study a $\mathcal{PT}$-symmetric two-level system and present its generalized IGP. The geometric phase transitions of the model at finite temperatures are located and analyzed.

The rest of the paper is organized as follows. Sec.~\ref{Sec2} briefly reviews the basics of PTQM and its statistical physics. We also review the geometric phases of pure and mixed quantum states in Hermitian systems via parallel-transport. In Sec. \ref{Sec3},  we generalize the formalism of geometric phase to PTQM, first by deriving two different expressions due to their associated evolution equations or parallel-transport conditions. We then generalize the results to thermal states in PTQM and derive the generalized IGP. Sec.~\ref{Sec4} presents the IGP of a $\mathcal{PT}$-symmetric two-level system and its geometric phase transitions at finite temperatures. Sec.~\ref{Sec5} concludes our work. Some details and derivations are summarized in the Appendix.


\section{Theoretical background}\label{Sec2}

\subsection{$\mathcal{PT}$-symmetric quantum and statistical mechanics}
Before presenting our findings, we first give a brief outline of PTQM and lay the foundation for its geometric description. We will set $c=\hbar=k_B=1$ throughout the paper.
We consider a parameter-dependent finite-dimensional non-Hermitian quantum system described by a $\mathcal{PT}$-symmetric Hamiltonian $H(\mathbf{R})$. Here $\mathbf{R}=(R_1,R_2,\cdots,R_k)^T$ is a collection of external parameters  forming a parameter manifold $M$. The system may evolve along a curve $\mathbf{R}(t)$ in $M$. The $\mathcal{PT}$-symmetry is manifested by the condition
\begin{align}\label{PT}
W(\mathbf{R})H(\mathbf{R})=H^\dag(\mathbf{R})W(\mathbf{R}),
\end{align}
where $W(\mathbf{R})$ is Hermitian, and its role will become clear later. A Hamiltonian satisfying Eq.~(\ref{PT}) is called a pseudo-Hermitian Hamiltonian \cite{PHQM11}.
Assuming $H$ describes a $N$-level quantum system,
the eigen-equations of $H(\mathbf{R})$ and $H^\dag(\mathbf{R})$ are respectively given by
\begin{align}
H(\mathbf{R})|\Psi_n(\mathbf{R})\rangle&=E_n(\mathbf{R})|\Psi_n(\mathbf{R})\rangle,\label{eve}\\
H^\dag(\mathbf{R})|\Phi_n(\mathbf{R})\rangle&=E_n(\mathbf{R})|\Phi_n(\mathbf{R})\rangle
\end{align}
for $n=1,2,\cdots N$. No energy degeneracy is considered here for simplicity. Eq.~(\ref{PT}) implies $|\Phi_n(\mathbf{R})\rangle=W(\mathbf{R})|\Psi_n(\mathbf{R})\rangle$. Here $W$ bears the role of a metric to ensure the orthonormal relation $\langle \Psi_m(\mathbf{R})|W(\mathbf{R})|\Psi_n(\mathbf{R})\rangle=\delta_{mn}$, or equivalently, $\langle \Phi_m(\mathbf{R})|\Psi_n(\mathbf{R})\rangle=\delta_{mn}$. Thus, the inner product between the ordinary bra and ket states is defined as $\langle \cdot |W|\cdot \rangle$. The associated completeness of $\{|\Psi_n(\mathbf{R})\rangle\}$ is given by $\sum_n|\Psi_n(\mathbf{R})\rangle\langle \Phi_n(\mathbf{R})|=1$.


Following Eq.~\eqref{PT}, $H$ is similar to a Hermitian Hamiltonian $H_0$ via $H=SH_0 S^{-1}$, where $W=(S^{-1})^\dag S^{-1}$~\cite{ZhangJMP20}. The operator $S$ is not unitary. Hereafter, we sometimes suppress the argument $\mathbf{R}$ if no confusion may arise. In some situations, $S$ may also be Hermitian and then $W=(S^{-1})^2$.
Diagonalizing $H_0$ as $H_0|\Psi^0_n\rangle=E_n|\Psi^0_n\rangle$, one gets
\begin{align}\label{WS}
|\Psi_n\rangle=S|\Psi^0_n\rangle,\quad |\Phi_n\rangle=(S^{-1})^\dag|\Psi^0_n\rangle.
\end{align}
For a generic time-dependent state $|\Psi(t)\rangle$ in PTQM, its equation of motion is described by the Schr\"odinger-like equation \cite{JPAGW13}:
 \begin{align}\label{eom1}
	\mi\frac{\dif}{\dif t}|\Psi(t)\rangle=\left(H-\frac{\mi}{2}  W^{-1}\dot{W}\right)|\Psi(t)\rangle.
\end{align}
If $S$ is a proper mapping satisfying $\dot{S}^{-1}S=(\dot{S}^{-1}S)^\dag$, this equation further reduces to
 \begin{align}\label{eom2}
	\mi\frac{\dif}{\dif t}|\Psi(t)\rangle=\tilde{H}|\Psi(t)\rangle.
\end{align}
Here $\tilde{H}=H-  S\dot{S}^{-1}$, and the second term is anti-symmetric under the $\mathcal{PT}$ transformation. Thus, $\tilde{H}$ is not Hermitian in general.
For PTQM, a proper $S$ always exists \cite{JPAGW13}.
Comparing Eqs.~(\ref{eom2}) and (\ref{eve}), it is important to emphasize that for a $\mathcal{PT}$-symmetric quantum system, the stationary and dynamic Schr\"odinger equations are respectively governed by $H$ and $\tilde{H}$. This distinction leads to nontrivial contributions to both the dynamic and geometric phases in PTQM, which will be elucidated in the subsequent discussions.
Introducing $|\Psi^0\rangle=S^{-1}|\Psi\rangle$, its dynamic evolution can be shown to obey the corresponding Schr\"odinger equation
 \begin{align}\label{eom3}
	\mi\frac{\dif}{\dif t}|\Psi^0(t)\rangle=H_0|\Psi^0(t)\rangle.
\end{align}
Thus, the proper $S$ acts like a ``gauge'' mapping between a PTQM system and its corresponding Hermitian counterpart.

So far the discussion concerns pure quantum states only.
Recently, there have been studies on non-Hermitian quantum models at finite temperatures \cite{PhysRevA.106.032206,NHQTD16,NHIsing23}.
To broaden the scope of non-Hermitian physics to mixed quantum states, we note that the density matrix of a mixed state from the generalization may also be non-Hermitian as well.
As a first attempt, we focus on states in thermal equilibrium depicted by $\rho=\frac{\me^{-\beta H}}{Z}$. Here $\beta=\frac{1}{T}$ is the inverse temperature and $Z=\sum_n\me^{-\beta E_n}$ is the partition function.
In the generalized case, $\rho^\dag\neq \rho$ due to $H^\dag\neq H$.
By expressing $H=\sum_n E_n|\Psi_n\rangle\langle \Phi_n|$, the density matrix is given by \begin{align}\label{rhoh}\rho=\sum_n\frac{\me^{-\beta E_n}}{Z}|\Psi_n\rangle\langle \Phi_n|,
\end{align}
whose trace follows the normalization $\text{Tr}\rho=\sum_n\langle \Phi_n|\rho|\Psi_n\rangle=1$. Applying Eq.~(\ref{WS}), we get a relation $\rho=S\rho_0S^{-1}$ connecting $\rho$ and $\rho_0=\frac{\me^{-\beta H_0}}{\text{Tr}\me^{-\beta H_0}}$.

\subsection{Geometric phase of Hermitian systems}
\subsubsection{Pure states}
The geometric phase, especially the Berry phase \cite{Berry84}, reflects the underlying geometry of quantum physics. For Hermitian systems, its formulation can be derived through the concept of the parallel condition among quantum states. Two states, $|\psi_{1}\rangle$ and  $|\psi_{2}\rangle$, are considered parallel with each other if $\langle \psi_1|\psi_2\rangle=\langle \psi_2|\psi_1\rangle>0$ \cite{WatrousBook}.
The overlap is also referred to as the fidelity~\cite{WatrousBook}. The parallel condition complements the concept of orthogonality of quantum states and builds a binary relation between quantum states. However, it is not an equivalence relation since it lacks transitivity. This means even when a state $|\Psi(t)\rangle\equiv|\Psi(\mathbf{R}(t))\rangle$ evolves along a path $\mathbf{R}(t)$ and preserves the condition of instantaneous parallel-transport, or being ``in-phase'', denoted as
 \begin{align}\label{pxc1}
 \langle \Psi(t)|\Psi(t+\dif t)\rangle>0,
\end{align}
it is possible that the final state may not remain parallel to the initial state. The loss of the parallelity is measured by the geometric phase, as explained here. By expanding the left-hand-side of Eq.~(\ref{pxc1}) and noticing that $\langle\Psi(t)|\frac{\dif}{\dif t}|\Psi(t)\rangle\dif t$ is imaginary, the parallel-transport condition is equivalent to \begin{align}\label{pxc2}
\langle\Psi(t)|\frac{\dif}{\dif t}|\Psi(t)\rangle=0.
\end{align}
The we rewrite $|\Psi(t)\rangle$ as $|\Psi(t)\rangle=\me^{\mi\theta(t)}|\psi(t)\rangle$, where $\theta(t)$ contains the information about the phase, including the dynamic and geometric components. However, the parallel-transport condition only allows the geometric phase to survive. Explicitly, if $|\Psi(t)\rangle$ experiences a dynamic evolution described by $\mi\frac{\dif}{\dif t}|\Psi(t)\rangle=H|\Psi(t)\rangle$ with $H$ being the Hamiltonian of a Hermitian quantum system, the condition (\ref{pxc2}) indicates $\mathlarger{\int_0^t}\dif t'\langle\Psi(t')|H|\Psi(t')\rangle=0$, i.e., the dynamic phase vanishes instantaneously. Substituting $|\Psi(t)\rangle=\me^{\mi\theta(t)}|\psi(t)\rangle$ into the parallel-transport condition, we get
 \begin{align}\label{BPen}
\mi\dot{\theta}+\langle\psi(t)|\frac{\dif}{\dif t}|\psi(t)\rangle=0.
\end{align}
In a cyclic process of duration $\tau$, the solution to Eq.~(\ref{BPen}) is the geometric phase
\begin{align}\label{BP}
\theta(\tau)=\mi\int_0^\tau\dif t\langle\psi(t)|\frac{\dif}{\dif t}|\psi(t)\rangle.
\end{align}

\subsubsection{Thermal states}
The geometric-phase formalism can be generalized to mixed quantum states undergoing a unitary evolution~\cite{GPMQS1}. When a density matrix evolves as $\rho(t)=U(t)\rho(0)U^\dag(t)$ with a unitary $U(t)$, it acquires a phase $\theta(t)=\arg\text{Tr}\left[\rho(0)U(t)\right]$.
Here ``Tr'' is the ordinary trace in the Hermitian quantum system.
It can be shown that $\rho(t+\dif t)
=U(t+\dif t)U^\dagger (t)\rho(t)U(t)U^\dagger(t+\dif t)$, yielding that $\rho(t)$ evolves into $\rho(t+\dif t)$ via $U(t+\dif t)U^\dagger (t)$. Accordingly, the condition $\arg\text{Tr}\left[\rho(t)U(t+\dif t)U^\dagger (t)\right]=0$ means that $\rho(t+\dif t)$ is ``in phase'' with $\rho(t)$ since no extra phase is accumulated during the evolution. Taking the differential form, we obtain the parallel-transport condition
\begin{align}\label{pxcm1}
\text{Tr}\left[\rho(t)\dot{U}(t)U^\dagger (t)\right]=\text{Tr}\left[\rho(0)U^\dag(t)\dot{U}(t)\right]=0.
\end{align}
Under this condition,
\begin{align}\label{GPm}
\theta_\text{G}(t)=\arg\text{Tr}\left[\rho(0)U(t)\right]
\end{align}
is the interferometric geometric phase (IGP), introduced in Ref.~\cite{GPMQS1}.
Similar to its pure-state counterpart, the parallel-transport condition (\ref{pxcm1}) also prevents the accumulation of the dynamic phase.
If $U(t)$ represents a dynamic evolution, then $\mi\dot{U}=HU$, or equivalently, $H=\mi\dot{U}U^\dag$. Thus, the dynamic phase accumulated during this evolution vanishes identically:
\begin{align}\label{Pdyn}
\theta_\text{D}(t)&=-\int_0^t\dif t'\text{Tr}\left[\rho(t')H(t')\right]\notag\\&=-\mi\int_0^t\dif t'\text{Tr}\left[\rho(t')\dot{U}(t')U^\dag(t')\right]=0.
\end{align}
If the trace is evaluated with the eigenstates $\{|n(t)\rangle\}$ of $\rho(t)$, only the diagonal elements $\langle n(t)|U(t)|n(t)\rangle$ is relevant to the determination of $\theta_\text{G}(t)$. Thus, to specify $U(t)$, it was suggested by  Sj$\ddot{\text{o}}$qvist et al. \cite{GPMQS1} to strengthen the parallel-transport condition as
 \begin{align}\label{pxcm2}
\langle n(t)|\dot{U}(t)U^\dag(t)|n(t)\rangle =0,\quad n=1,2,\cdots, N.
\end{align}
If $\rho(t)$ is the density matrix of a pure state, Eq.~(\ref{pxcm1}) naturally reduces to the condition (\ref{pxc2}) for pure states.

\section{Geometric phase of PT-symmetric quantum systems}\label{Sec3}
\subsection{Geometric phase for pure states}
\subsubsection{Adiabatic approaches}
The concept of geometric phase has been generalized to some non-Hermitian systems in Ref. \cite{JPAGW13}, where the expression of the Berry phase was obtained by following Berry's formalism of adiabatic evolution. Explicitly, for a $\mathcal{PT}$-symmetric system undergoing evolution along a loop $C(t):=\mathbf{R}(t)$ with $0<t<\tau$ and $\mathbf{R}(0)=\mathbf{R}(\tau)$ in the parameter manifold, the $n$th eigenstate at the end of this evolution is given by
 \begin{align}\label{thetatot}|\Psi_n(\mathbf{R}(\tau))\rangle=\me^{\mi\theta^\text{D}_{n}(\tau)+\mi\theta^\text{B}_n(C)}|\Psi_n(\mathbf{R}(0))\rangle.\end{align}
Here, $\theta^\text{D}_{n}(t)=- \mathlarger{\int}_0^t \mathrm{d} t'E_n(\mathbf{R}(t'))$ represents the instantaneous dynamic phase, and
 \begin{align}\label{thetaBn1}
 \theta^\text{B}_n(C)=\mi\oint_C\dif \mathbf{R}\cdot\left[\left\langle\Psi_n|W \nabla| \Psi_n\right\rangle+\frac{1}{2}\left\langle\Psi_n|(\nabla W)| \Psi_n\right\rangle\right]
\end{align}
is the Berry phase of PTQM following this approach.
It should be noted that this result is obtained by beginning with the stationary Schr\"odinger equation shown in Eq.~(\ref{eve}) \cite{JPAGW13}. In this approach, $\theta^\text{D}_{n}(t)$ is generated through the time evolution controlled by $H_0$, as indicated by Eq.~(\ref{eom3}).

Meanwhile, a different approach is based on the time evolution described by Eq.~(\ref{eom2}), whose dynamics is governed by the effective Hamiltonian $\tilde{H}=H-\mi S\dot{S}^{-1}$. Different from the prior approach, it will be shown that the ``gauge'' map $S$ imparts significant effects on both the dynamic and geometric phases. This also influences the generalization of the geometric phase to thermal states in $\mathcal{PT}$-symmetric systems.

When following Eq.~(\ref{eom2}) along the loop $C(t)$, the $n$th eigenstate acquires an instantaneous dynamic phase
\begin{align} \label{DPn}
\theta^1_{\text{D}n}(t)&=- \int_0^t \mathrm{d} t'\langle\Psi_n(t')|W\tilde{H}| \Psi_n(t')\rangle\notag\\
&=- \int_0^t \mathrm{d} t'E_n(t')+\mi\int_0^t \mathrm{d} t'\langle \Psi^0_n(t')|\dot{S}^{-1}S|\Psi^0_n(t')\rangle\notag\\
&=\theta^\text{D}_{n}(t)-\mi\int_0^t \mathrm{d} t'\langle \Psi^0_n(t')|S^{-1}\dot{S}|\Psi^0_n(t')\rangle,
\end{align}
where $|\Psi_n(t)\rangle\equiv |\Psi_n(\mathbf{R}(t))\rangle$ and $|\Psi^0_n(t)\rangle\equiv |\Psi_n^0(\mathbf{R}(t))\rangle$.
Importantly, $\theta^\text{D}_{n}(t)$ is real-valued, while $\theta^1_{\text{D}n}(t)$ is in general complex-valued since the dynamic equation (\ref{eom2}) is governed by the non-Hermitian $\tilde{H}$. This is reasonable since PTQM may be realized by open systems, and complex phases implies gain or decay of the amplitude. Moreover, the second term in the last line of Eq.~(\ref{DPn}) is purely imaginary if $S$ is a proper mapping. To derive the geometric phase, we consider a state $|\Psi(t)\rangle$ and expand it in terms of the instantaneous eigenstates of $H(t)$ as
\begin{align} \label{Pt}
|\Psi(t)\rangle=\sum_nc_n(t)\me^{\mi \theta^1_{\text{D}n}(t)}|\Psi_n(t)\rangle.
\end{align}
If the system experiences an adiabatic evolution along $C(t)$, no level crossing occurs. Thus, we found $c_n(t)\approx c_n(0)\me^{\mi\theta^1_n(t)}$, or
\begin{align} \label{Ct} |\Psi_n(t)\rangle&=\me^{\mi \theta^1_{\text{D}n}(t)+\mi\theta^1_n(t)}|\Psi_n(0)\rangle.
\end{align}
Here
\begin{align} \label{theta1t}
\theta^1_n(t)&=\mi\int_0^t\dif t'\langle \Phi_n(t')|\frac{\dif}{\dif t'}|\Psi_n(t')\rangle.
\end{align}
A detailed derivation is outlined in Appendix \ref{appa}.
We come to an interesting result: There exist two types of geometric phases in PTQM due to the  evolutionary equations associated with the non-Hermitian Hamiltonian and its Hermitian counterpart.

\subsubsection{Parallel-transport conditions}
What is the relation between the geometric phases derived previously? Moreover, we have pointed out that there is an equivalent way to derive the geometric phase based on the parallelity between quantum states in conventional QM.
Does this approach also apply to PTQM?
To answer these questions, we first generalize the previously introduced parallel-transport condition to PTQM. Note the time evolution (\ref{eom2}) in a $\mathcal{PT}$-symmetric system is controlled by $H$, which is related to the Hermitian Hamiltonian $H_0$ that governs the dynamic equation (\ref{eom3}) via a similarity transformation $S$.

It has been shown that in conventional QM, the parallel-transport condition (\ref{pxc2}) ensures that the dynamic phase vanishes. Equivalently, the appearance of a non-vanishing dynamic phase violates the instantaneous  parallelity when a state is evolved. Hence, in order to avoid violation of the instantaneous parallelity,
we follow an approach similar to that of conventional QM to remove the dynamic phase $\theta^1_{\text{D}n}$ from Eq.~(\ref{Pt}) and introduce $|\tilde{\Psi}_n(t)\rangle=\me^{\mi\theta^1_n(t)}|\Psi_n(t)\rangle$. Similarly, we also define $|\tilde{\Psi}^0_n(t)\rangle=\me^{\mi\theta^2_n(t)}|\Psi^0_n(t)\rangle$ by eliminating $\theta^\text{D}_{n}$ generated during a dynamic evolution controlled by $H_0$.
A generalizations of Eq.~(\ref{pxc2}) leads to the following parallel-transport (or instantaneous in-phase) conditions:
\begin{align}
\langle\tilde{\Phi}_n(t)|\frac{\dif}{\dif t}|\tilde{\Psi}_n(t)\rangle&=0,\label{pxc3a} \\
\langle\tilde{\Psi}^0_n(t)|\frac{\dif}{\dif t}|\tilde{\Psi}^0_n(t)\rangle&=0.\label{pxc3b}
\end{align}
Thus, $\theta^{1,2}_n(t)$ is the accumulated phase during the respective parallel transport.
Solving these equations, we get
\begin{align}
\theta^1_n(C)&=\mi\oint_C\dif t\langle \Phi_n(t)|\frac{\dif}{\dif t}|\Psi_n(t)\rangle,\label{thetaBn2}\\
\theta^2_n(C)&=\mi\oint_C\dif t\langle \Psi^0_n(t)|\frac{\dif}{\dif t}|\Psi^0_n(t)\rangle,\label{thetaBn2b}
\end{align}
at the end of the corresponding parallel transport. Eq.~(\ref{pxc3a}) reproduces the geometric phase of Eq.~(\ref{theta1t}) derived by the adiabatic approach. Moreover, it can be verified that $\theta^2_n$ matches the Berry phase shown in Eq.~\eqref{thetaBn1}:
\begin{align}\label{theta1B}\theta^2_n(C)=\theta^\text{B}_n(C)\end{align}
as long as $S$ is a proper mapping.

While Eq.~(\ref{DPn}) gives a relation between the two dynamic phases $\theta^1_{\text{D}n}$ and $\theta^\text{D}_n$,
there is a similar relation connecting $\theta^{1}_n$ and $\theta^{2}_n$:
\begin{align} \label{Berry0}
&\theta^1_n=\theta^2_n+\mi \oint\dif t\langle\Psi^0_n(t)|S^{-1} \dot{S}| \Psi^0_n(t)\rangle.
\end{align}
The proofs of Eqs.~(\ref{theta1B}) and (\ref{Berry0}) are outlined in Appendix \ref{appa}.
Interestingly, $\theta^{1}_n$ can be complex-valued due to the presence of the non-Hermitian $\tilde{H}$ in the dynamic evolution (\ref{eom2}). Since the dynamic phase  $\theta^1_{\text{D}n}$ is excluded by the parallel-transport condition, what remains is the geometric component $\theta^1_{n}$. The imaginary part of $\theta^1_{n}$ implies a change of the amplitude of the wavefunction since the system is non-Hermitian. We will find similar results in our subsequent discussions on thermal states.


In the framework of the IGP, the geometric phase for mixed states is intricately linked to that of pure states. This raises a pertinent question: In the context of PTQM, which one of $\theta^{1,2}_n$ is more natural for a generalization to thermal states? Referring back to Eqs.~(\ref{eom3}), (\ref{thetatot}), and (\ref{pxc3b}), it can be inferred that both $\theta^2_n$ and $\theta^\text{D}_n$ may arise in a quantum system governed by $H_0$. In contrast, $\theta^1_n$ and $\theta^1_{\text{D}n}$ can be generated in a $\mathcal{PT}$-symmetric system controlled by $H$. Consequently, we choose $\theta^1_n$ and the corresponding approach to develop the formalism of the IGP of thermal states in PTQM.

\subsection{Interferometric geometric phase for thermal states}
To generalize the IGP to PTQM, we focus on states in thermal equilibrium at temperature $T$ described by their non-Hermitian density matrix $\rho=\frac{1}{Z}\me^{-\beta H}$ as stated before.
Since the density matrix may be a non-Hermitian operator in those cases, it usually experiences non-unitary evolution since $H$ is non-Hermitian. We consider a general form
$\rho(t)=U(t)\rho(0)U^{-1}(t)$ with $\rho(0)=\rho$. Similar to conventional QM, the system acquires a (total) phase \begin{align}\label{phtot}\theta_\text{tot}(t)=\arg\text{Tr}\left[\rho(0)U(t)\right]\end{align} during this evolution.
Since a statistical ensemble encompasses all energy levels, each weighted by its respective thermal weight, it is more suitable to introduce the geometric phase via the parallel-transport condition, which also fixes the form of $U(t)$.
To ensure that $\rho(t+\dif t)$ is in phase with $\rho(t)$ during the evolution, the condition (\ref{pxcm1}) is generalized as
\begin{align}\label{pxcmnh0}
\text{Tr}\left[\rho(t)\dot{U}(t)U^{-1}(t)\right]=\text{Tr}\left[\rho(0)U^{-1}(t)\dot{U}(t)\right]=0.
\end{align}
If $U(t)$ is a time evolution along a loop in the parameter manifold, then Eq.~(\ref{eom2}) yields $\mi\dot{U}=\tilde{H}U$ or $\mi\dot{U}U^{-1}=\tilde{H}$.
Similar to Eq.~(\ref{Pdyn}), the parallel-transport condition (\ref{pxcmnh0}) causes the dynamic phase to vanish:
\begin{align}
\theta_\text{D}(t)=\int_0^t\dif t'\text{Tr}\left[\rho(t')\tilde{H}(t')\right]=0.
\end{align}
This may be realized by choosing a suitable evolution path in the parameter manifold \cite{Hou2023}. Under parallel-transport, the density matrix evolves as
\begin{align}\label{rhoht}\rho(t)=\sum_n\frac{\me^{-\beta E_n}}{Z}|\Psi_n(t)\rangle\langle \Phi_n(t)|,
\end{align}
where $|\Psi_n(t)\rangle \equiv|\Psi_n(\mathbf{R}(t))\rangle $ and $|\Phi_n(t)\rangle \equiv|\Phi_n(\mathbf{R}(t))\rangle $. The trace in Eq.~(\ref{pxcmnh0}) can be evaluated as $\sum_n\langle\Phi_n(t)|\cdot|\Psi_n(t)\rangle$. Similar to Eq.~(\ref{pxcm2}), the parallel-transport condition are also reinforced as
 \begin{align}\label{pxcmnh1}
\langle \Phi_n(t)|\dot{U}(t)U^{-1}(t)|\Psi_n(t)\rangle =0,\quad n=1,2,\cdots, N.
\end{align}
Since the dynamic phase vanishes during parallel-transport, the system acquires the IGP according to Eq.~(\ref{phtot}):
 \begin{align}\label{GPmnh}
\theta_\text{G}(t)=\theta_\text{tot}(t)=\arg\text{Tr}\left[\rho(0)U(t)\right].
\end{align}

A transformation satisfying the parallel-transport condition has the form
\begin{align}\label{Ut}
	U(t)=&\sum_n\me^{- \int^{t}_0  \langle\Phi_n(t') |\frac{\dif}{\dif t'}| \Psi_n(t') \rangle \dif t'}|\Psi_n( t )\rangle \langle\Phi_n(0)|\notag\\
=&\sum_n\me^{\mi \theta^1_n(t)}|\Psi_n( t )\rangle \langle\Phi_n(0)|.
\end{align}
Similar to its pure-state counterpart, the dynamic phase $\theta^1_{\text{D}n}(t)$ for each level is not included to avoid violating the parallel-transport condition.
Appendix \ref{appb} shows how $U(t)$ indeed satisfies the condition (\ref{pxcmnh1}). The IGP accumulated during the evolution is
\begin{align}\label{IGPt}
\theta_\text{G}(t)=\arg\left[\sum_n\frac{\me^{-\beta E_n}}{Z}\me^{- \int^{t}_0  \langle\Phi_n(t') |\frac{\dif}{\dif t'}| \Psi_n(t') \rangle \dif t'}\nu_n(t)\right],
\end{align}
where $\nu_n(t)=\langle\Phi_n(0)|\Psi_n( t )\rangle$.
If the system undergoes a cyclic process along a loop $C(t)=\mathbf{R}(t)$ with $\mathbf{R}(\tau)=\mathbf{R}(0)$, then $\nu_n(\tau)=1$ and
\begin{align}\label{IGPC}
\theta_\text{G}(C)=\arg\left[\sum_n\frac{\me^{-\beta E_n}}{Z}\me^{\mi\theta^1_n(C)}\right].
\end{align}
Here $\theta^1_n(C)$ is the geometric phase factor associated
with the $n$th individual pure state in the process, given by Eq.~(\ref{thetaBn2}). It can be shown that $\theta_\text{G}(C)$ reduces to $\theta^1_n(C)$ in the zero temperature limit since $\lim_{\beta\rightarrow \infty}\frac{\me^{-\beta E_1}}{Z}=1$ and $\lim_{\beta\rightarrow \infty}\frac{\me^{-\beta E_{n>1}}}{Z}=0$. This is consistent with the reason that we choose $\theta^1_n(C)$ as the geometric phase for pure states in PTQM. Since $\theta^1_n(C)$ is complex in general, $\theta_\text{G}(C)$ may also be complex. Its effect will be clarified later.

\section{Example: two-level system}\label{Sec4}
To understand the IGP of PTQM more clearly, we study a $\mathcal{PT}$-symmetric two-level system introduced in Refs.~\cite{Wang_2010,JPAGW13} and calculate its IGP. The Hamiltonian is given by
\begin{align}\label{Hberry}
	H=\epsilon \mathbf{1}_{2 \times 2}+\left(a \mathbf{n}^r+\mathrm{i} b  \mathbf{n}^\theta\right) \cdot \boldsymbol{\sigma},
\end{align}
where $\boldsymbol{\sigma}=(\sigma_x,\sigma_y,\sigma_z)^T$ is the collection of Pauli matrices and $\mathbf{n}^r \equiv(\sin \theta \cos \phi, \sin \theta \sin \phi, \cos \theta)^T$, $\mathbf{n}^\theta \equiv(\cos \theta \cos \phi, \cos \theta \sin \phi,-\sin \theta)^T$ are the unit vectors respectively along the radial and tangent directions of a meridian on a unit sphere. The eigenvalues are $E_{\pm}=\epsilon\pm \sqrt{a^2-b^2}$.
We limit our discussion to the regime of $a^2>b^2$, where the $\mathcal{PT}$-symmetry is not broken and $E_\pm$ is real.
Without loss of generality, we let $a>0$. The two eigen-vectors are
\begin{align}\label{HFeigenv}
		& \left|\Psi_{+}\right\rangle=n_{+}\left(\begin{array}{c}
			\left(\cos \frac{\theta}{2}-\mi\alpha  \sin \frac{\theta}{2}\right) \mathrm{e}^{-\mathrm{i} \phi} \\
			\mi\alpha  \cos \frac{\theta}{2}+\sin \frac{\theta}{2}
		\end{array}\right), \\
		& \left|\Psi_{-}\right\rangle=n_{-}\left(\begin{array}{c}
			-\left(\mi\alpha  \cos \frac{\theta}{2}+\sin \frac{\theta}{2}\right) \mathrm{e}^{-\mathrm{i} \phi} \\
			\cos \frac{\theta}{2}-\mi\alpha  \sin \frac{\theta}{2}
		\end{array}\right),
\end{align}
where $\alpha = \frac{b}{a+\sqrt{a^2-b^2}}$ and $n_{ \pm}=\mathrm{e}^{-\mathrm{i} \frac{\theta}{2}} \sqrt{\frac{a^2+a \sqrt{a^2-b^2}}{2\left(a^2-b^2\right)}}$ are normalization coefficients. The metric operator $W$ of this case is
 \begin{align}\label{W}
W=1-\frac{b}{a}\mathbf{n}^\phi \cdot \boldsymbol{\sigma},
 \end{align}
where $\mathbf{n}^\phi=(-\sin\phi,\cos\phi,0)^T$ is the unit tangent vector of a latitude.
In what follows, we will fix $a$ and $b$, thus the parameters $(\theta,\phi)$ form the parameter manifold $S^2$, a unit spherical surface.
Using Eq.~(\ref{thetaBn1}), $\theta^2_{ \pm}$ associated with a loop $C$ on $S^2$ is given by \cite{JPAGW13}
 \begin{align}\label{theta2n}
 \theta^2_{ \pm}(C)=&\mp \frac{1}{2} \frac{a}{\sqrt{a^2-b^2}} \Omega(C)+\left(1 \pm \frac{a}{\sqrt{a^2-b^2}}\right) \pi \notag\\ \pm& \frac{\pi}{4}\left(1-\frac{a}{\sqrt{a^2-b^2}}\right),
  \end{align}
if the north pole is enclosed by $C$, or
 \begin{align}\label{gppm}
 \theta^2_{ \pm}(C)=\mp \frac{1}{2} \frac{a}{\sqrt{a^2-b^2}} \Omega(C) \pm \frac{\pi}{4}\left(1-\frac{a}{\sqrt{a^2-b^2}}\right)
 \end{align}
 if the north pole is not enclosed by $C$. Here $\Omega(C)=\mathlarger{\oint}_C\dif\phi(1-\cos\theta)$ is the solid angle of the surface enclosed by $C$ with respect to the origin.

As a concrete example, we take $a=3$ and $b=\sqrt{5}$, so the eigenvalues become $E_\pm=\epsilon\pm 2$.
To calculate the second term of $\theta^1_\pm(C)$ in Eq.~(\ref{thetaBn2b}), we choose a proper $S$, which can be constructed via the procedure of Ref. \cite{JPAGW13} summarized in Appendix~\ref{appc}. Explicitly, it is given by
 \begin{align}
S_{\text {proper }}=
	\begin{pmatrix}
		\frac{1}{2} \sqrt{\frac{15}{2}}  \me^{\frac{ \mi \phi }{4}} & -\frac{1}{2}  \mi \sqrt{\frac{3}{2}}  \me^{-\frac{5  \mi \phi}{4} } \\
		\frac{1}{2} \mi \sqrt{\frac{3}{2}}  \me^{\frac{5  \mi \phi }{4}} & \frac{1}{2} \sqrt{\frac{15}{2}}  \me^{-\frac{\mi \phi}{4} }
	\end{pmatrix}.
 \end{align}
 Under this proper transformation, the original non-Hermitian Hamiltonian is converted into a Hermitian one:
 \begin{align}
	H_0=S^{-1}_{\text {proper }} H S_{\text {proper }}=\begin{pmatrix}
		\epsilon +2 \cos \theta  & 2 \me^{-\frac{3 \mi \phi }{2} } \sin \theta  \\
		2 \me^{\frac{3 \mi \phi }{2}} \sin \theta  & \epsilon -2 \cos \theta
		\end{pmatrix}.
\end{align}
The eigenvector associated with $\epsilon$+2 is
\begin{align}
	\left| \Psi^0_+(\theta,\phi)\right\rangle=\begin{pmatrix}
		\frac{\me^{-\frac{3 \mi \phi }{2} } (\cot \theta +\csc \theta )}{\sqrt{(\cot \theta +\csc \theta )^2+1}}\\
		\frac{1}{\sqrt{(\cot \theta +\csc \theta )^2+1}}
	\end{pmatrix}.
\end{align}
It can be shown that
\begin{align}
	\langle \Psi^0_+|  S^{-1} \dif S	| \Psi^0_+\rangle&=-\frac{1}{4} \sqrt{5} \sin \theta \dif \phi, \notag\\
	\mi \oint_C \langle\Psi^0_+ |S^{-1} \dot{S} | \Psi^0_+ \rangle \dif t&=-\mi\frac{\pi}{2}\sqrt{5} \sin \theta ,  \label{phase2}
\end{align}
where the loop $C$ is chosen as a circle of latitude $\theta$.
Similarly, the imaginary part of $\theta^1_-$ is
\begin{align}
	\mi \oint_C\langle\Psi^0_- |S^{-1} \dot{S} | \Psi^0_- \rangle \dif t=\mi\frac{\pi}{2}\sqrt{5} \sin \theta.  \label{phase3}
\end{align}
Since the north pole is enclosed by $C$ (a circle of latitude), $\theta^2_\pm(C)$ is evaluated by Eq.~(\ref{phase2}).
Using Eqs.~(\ref{Berry0}), (\ref{phase2}), and (\ref{phase3}), the geometric phases associated with the two eigenstates are
\begin{align}\label{thetag1}
\theta^1_+(C)	&=-\frac{3\pi}{2}(1-\cos \theta )+\frac{3\pi}{8}-\mi\frac{\pi}{2}\sqrt{5} \sin \theta ,\notag\\
\theta^1_-(C)	&=\frac{3\pi}{2}(1-\cos \theta )-\frac{3\pi}{8}+\mi\frac{\pi}{2}\sqrt{5} \sin \theta,
\end{align}
respectively.
Here an extra factor $2\pi$ is dropped from $\theta^1_+(C)$.
Accordingly, the IGP is
\begin{align}\label{final1}
&	\theta_\text{G} (C) 	=\arg \left[\frac{\me^{-2\beta}\me^{\mi\theta^1_+(C)}+ \me^{2\beta}\me^{\mi\theta^1_-(C)} }{\me^{2\beta}+\me^{-2\beta}  }  \right]\notag\\
&=\arg \left[\me^{-2\beta+\frac{\sqrt{5}\pi}{2}\sin\theta}\me^{\mi\theta^2_+(C)}+ \me^{2\beta-\frac{\sqrt{5}\pi}{2}\sin\theta}\me^{\mi\theta^2_-(C)} \right],
\end{align}
where $\theta^{2}_\pm$ is the real part of $\theta^{1}_\pm$, as shown by Eq.~(\ref{Berry0}). The imaginary part of $\theta^{1}_\pm$ actually changes the thermal weight of each energy-level.

 \begin{figure}[t]
\centering
\includegraphics[width=3.2in,clip]{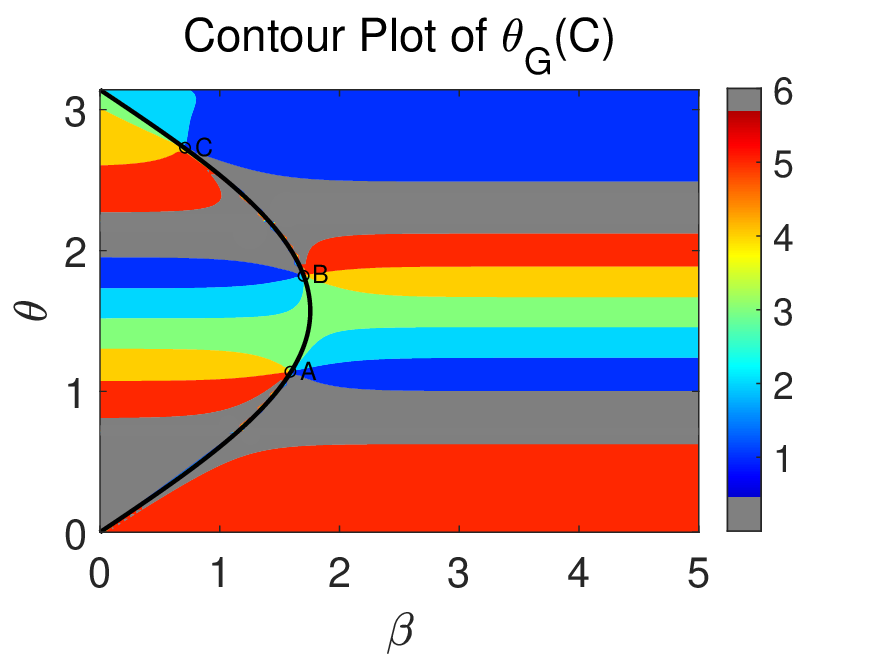}
\caption{Contour plot of $\theta_\text{G}(C)$ as a function of $\beta$ and $\theta \in [0,\pi]$, where the range of $\theta_\text{G}(C)$ is within $[0,2\pi]$. The black curve shows the arc $\beta=\frac{\sqrt{5}\pi \sin\theta}{4}$, and the value of $\theta_\text{G}(C)$ jumps at the singular points A, B and C. }
\label{Fig1}
\end{figure}

\begin{figure}[t]
\centering
\includegraphics[width=3.0in,clip]{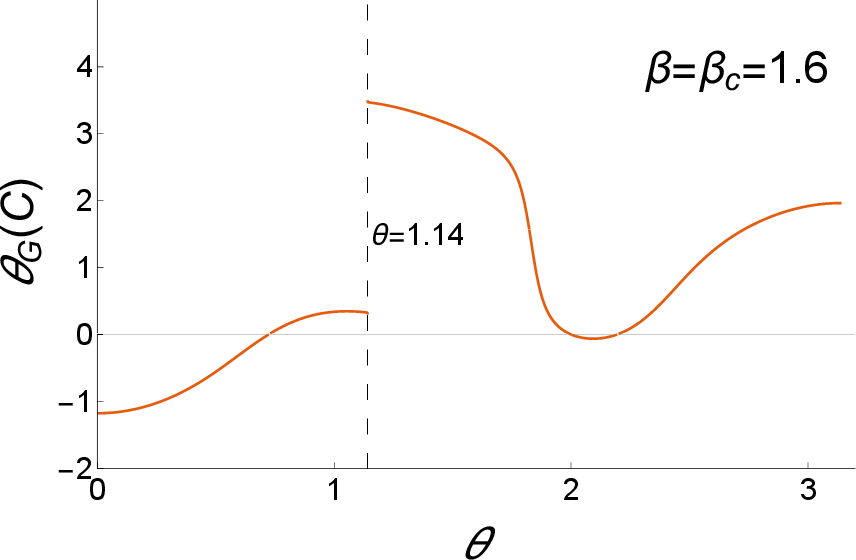}
\includegraphics[width=3.0in,clip]{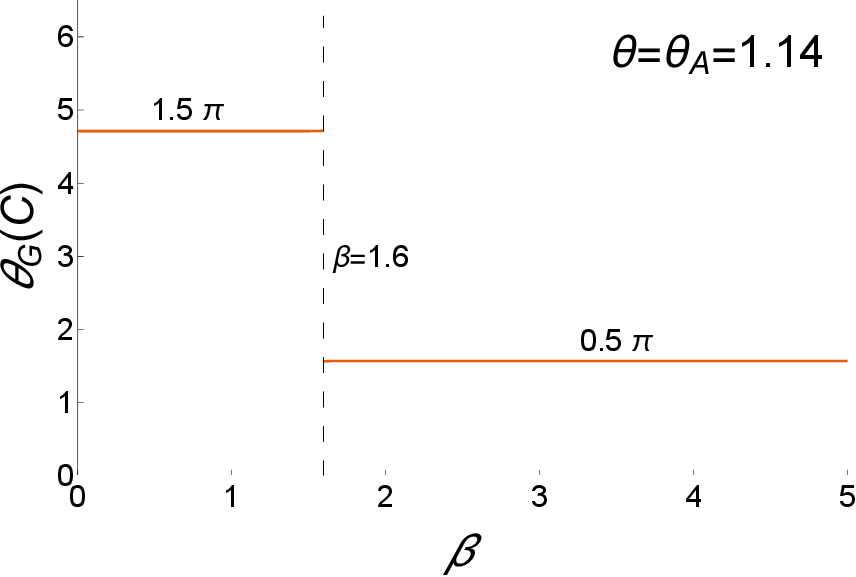}
\caption{(Top panel) $\theta_\text{G}(C)$ as a function of $\theta$ at fixed  $T=\frac{1}{\beta_c}$. When crossing the singular point $\theta_\text{A}\approx 1.14$, there is a $\pi$-jump in $\theta_\text{G}(C)$. (Bottom panel) $\theta_\text{G}(C)$ as a function of $\beta$ for the evolution along the circle of latitude with $\theta=1.14$. As the system crosses the critical inverse temperature $\beta_c=1.6$, $\theta_\text{G}(C)$ exhibits a $\pi$-jump. }
\label{Fig2}
\end{figure}

\begin{figure}[t]
\centering
\includegraphics[width=3.2in,clip]{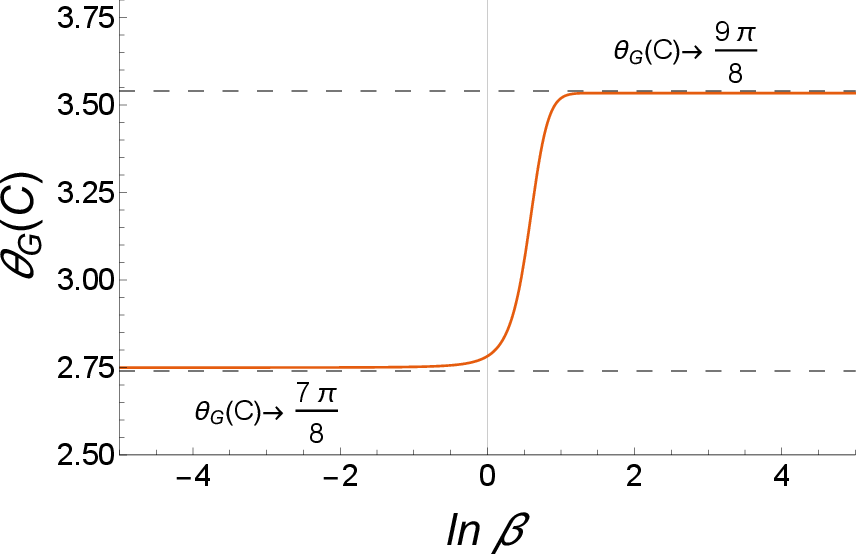}
\caption{ $\theta_\text{G}(C)$ as a function of $\ln\beta$ when the system evolves along the equator ($\theta=\frac{\pi}{2}$). In this case, the imaginary part of $\theta^1_\pm$ has a maximal effect on the thermal weights. }
\label{Fig3}
\end{figure}

Eq.~(\ref{GPm}) shows that the IGP is the argument of $\text{Tr}\left[\rho(0)U(t)\right]$, which is the ``returning amplitude'' between the initial state $\rho(0)$ and the instantaneous state $\rho(t)$ \cite{GPMQS1,Hou2023}. It can also be thought of as a generalization of the Loschmidt amplitude in mixed quantum states. At its zeros, the IGP exhibits discontinuities and nonanalytical behavior, signaling a change of the geometric nature of the system reflected by the IGP. In this example, the second line of Eq.~(\ref{final1}) shows that $\theta_\text{G}(C)$ may become singular if $\beta=\frac{\sqrt{5}\pi \sin\theta}{4}$. To examine the IGP of PTQM, we visualize our findings in Figs.~\ref{Fig1}, \ref{Fig2}, and \ref{Fig3}.

In Fig.~\ref{Fig1}, we present the contour plot of $\theta_\text{G}(C)$ as a function of $\beta$ and $\theta$. Indeed, there are three singular points A, B and C lying on the arc $\beta=\frac{\sqrt{5}\pi \sin\theta}{4}$, which correspond to the latitudes $\theta_{\text{A,B,C}}=\arccos\left(\frac{5}{12}\right)\approx 1.14$, $\arccos\left(-\frac{1}{4}\right)\approx1.82$, and $\arccos\left(-\frac{11}{12}\right)\approx2.73$, respectively. The IGP changes rapidly near A, B and C, indicating that the value of $\theta_\text{G}(C)$ jumps discretely when crossing these singular points. Notably, a jump of the IGP at finite temperature has been ruled out in any two-level model of Hermitian quantum systems \cite{Hou2023}.

To grasp the physical significance of the arc $\beta=\frac{\sqrt{5}\pi \sin\theta}{4}$, we revisit the corresponding Hermitian quantum system, where the thermal weight of each level is proportional to $\me^{\mp 2\beta}$ at temperature $T=\frac{1}{\beta}$. As $T\rightarrow 0$, the relative weight between the excited and ground states becomes $\lim_{\beta\rightarrow +\infty}\frac{\me^{-2\beta}}{\me^{2\beta}}=0$, leading the IGP to converge to the geometric phase of the ground state. In the infinite temperature limit ($\beta\rightarrow 0$), the relative weight becomes $\lim_{\beta\rightarrow 0}\frac{\me^{-2\beta}}{\me^{2\beta}}=1$. In this case, the Hermitian density matrix corresponds to the maximally mixed state, where each level has equal thermal weight, and the IGP loses its resemblance to the ground-state geometric phase. Turning to $\mathcal{PT}$-symmetric systems, the parallel-transport condition eliminates the dynamic phase from the total phase, leaving a complex IGP. The imaginary part of the IGP (or $\theta^1_\pm$) modifies the thermal weights of the two levels to exp$\left[\mp\left(2\beta-\frac{\sqrt{5}\pi \sin\theta}{2}\right)\right]$, which will be referred to as the ``effective thermal weights''. Notably, in the low-temperature limit, the behavior of the IGP can still mirror that of the corresponding Hermitian system. In Fig.~\ref{Fig1},
the domain where $\beta>\frac{\sqrt{5}\pi \sin\theta}{4}$ corresponds to the phase at ``effective'' positive temperatures for the non-Hermitian quantum system. The arc $\beta=\frac{\sqrt{5}\pi \sin\theta}{4}$ signifies the ``effective'' infinite-temperature threshold. Conversely, the regime where $\beta<\frac{\sqrt{5}\pi \sin\theta}{4}$ corresponds to the phase at ``effective'' negative temperatures.
In this scenario, the original temperature $T$ alongside with the imaginary part of $\theta^1_\pm$ determines the relative thermal distribution between the excited and ground states.

At the singular points A, B and C on the curve $\beta=\frac{\sqrt{5}\pi \sin\theta}{4}$, it can be verified that $\theta^2_-(C)-\theta^2_+(C)=\pi$, $3\pi$ and $5\pi$. Thus, the genuine geometric phase factor $\me^{\mi\theta^2_\pm(C)}$ associated with each level is off by a factor of $-1$, making the IGP jump by $\pi$ when crossing these points. The physical meaning behind these discontinuities can be understood as follows. Take for example point A with critical inverse temperature $\beta_c=1.6$. When the system at the fixed temperature $T_c=\frac{1}{\beta_c}$ evolves along a circles of latitude $\theta=\theta_\text{A} + 0^+$ (or $\theta=\theta_\text{A} - 0^+$), the system ends up in the ``effective'' positive (or negative) temperature phase. When crossing $\theta_\text{A}$, the IGP experiences a $\pi$-jump.
To better visualize the phenomena, we show the jump of the IGP along $\beta_c=1.6$ in the top panel of Fig.~\ref{Fig2}. Similarly, the geometric nature of the evolution along a circle of latitude $\theta=\theta_\text{A}$ changes as the inverse temperature crosses $\beta_c$.  This transition is clearly depicted in the bottom panel of Fig.~\ref{Fig2}. We refer to this non-analytical behavior of the IGP as signaling a geometric phase transition. Explicitly, the system at point A is in the ``effective'' positive-temperature phase when $\beta>\beta_c$ and the IGP resembles $\theta^2_-(C)=\frac{\pi}{2}$, the real-valued geometric phase of the ground state. After crossing $\beta_c$, the system enters the ``effective'' negative-temperature phase with $\beta<\beta_c$, and the IGP changes to resemble $\theta^2_+(C)=-\frac{\pi}{2}\equiv \frac{3\pi}{2}\mod 2\pi$, the real-valued geometric phase of the excited state.

When $\theta\neq \theta_\text{A,B,C}$, the IGP represents an interpolation between the geometric phases of the excited and ground states as temperature varies. When the system evolves along the equator with $\theta=\frac{\pi}{2}$, the magnitude of $|\text{Im}\theta^1_\pm(C)|$ reaches its maximum, indicating a maximal contribution to the effective temperature in determining the thermal weights.
Furthermore, $\theta^2_\pm(C)=\text{Re}\theta^1_\pm(C)=\mp \frac{9\pi}{8}$, as derived from Eq.~(\ref{thetag1}).
The behavior of $\theta_\text{G}(C)$ is illustrated in Fig.~\ref{Fig3}, where $\beta$ transitions from $0$ (the infinite-temperature limit) to $+\infty$ (the low-temperature limit) displayed on a logarithmic scale. In the scenario where $\beta\rightarrow +\infty$, $\theta_\text{G}(C)\rightarrow \frac{9\pi}{8}=\theta^2_-(C)$. Conversely, when $\beta$ approaches 0, $\theta_\text{G}(C)$ approximates $\theta^2_+(C)$, which equals $\frac{7\pi}{8}\equiv -\frac{9\pi}{8}\mod 2\pi$.

\subsection{Implications}
On the one hand, PTQM may be realized in driven systems. For example, Ref.~\cite{R_ter_2010} demonstrated a $\mathcal{PT}$-symmetric quantum system with two coupled optical waveguides selectively pumped. 
By modulating the refractive index along the waveguides, the Hamiltonian may be engineered to the desired form.
On the other hand, the IGP of mixed states in Hermitian systems have been  measured by using a Mach-Zehnder interferometer setup demonstrated in Refs.~\cite{Du03,Ericsson05}, where mixed states were generated through two methods: Decohering pure states with birefringent elements and creating a non-maximally entangled state of two photons followed by tracing out one photon.

As shown in this work, the IGP of PTQM is in general complex-valued, where the real part represents a phase factor while the imaginary part adjusts the distribution. By applying the phase measurement~\cite{Du03,Ericsson05} to extract the IGP of mixed states in $\mathcal{PT}$-symmetric systems, it is likely to extract only the thermal average of the IGPs of individual states. Nevertheless, one may compare the population distribution of the evolved system with that of a corresponding system without the accumulation of the IGP. The difference in the distribution is due to the imaginary part of the IGP of the PTQM system. Therefore, the real- and imaginary- parts of the IGP of PTQM systems seem to be measurable albeit the procedure is more complicated due to the lack of Hermiticity.

\section{Conclusion}\label{Sec5}
The concept of geometric phase has been generalized to  PTQM via the introduction of parallel-transport. For pure-states, the parallel-transport conditions for the eigenstates of $H$ and $H_0$ lead to distinct generalizations of the geometric phases, $\theta^1$ and $\theta^2$, as obtained from the conventional methods. In general, $\theta^1$ is complex and $\theta^2$ is its real part. As  $\theta^1$ arises from the non-Hermitian Hamiltonian, it is generalized to mixed states in PTQM. Moreover, the discussion of the IGP of mixed states is meaningful after the dynamic phase has been removed by the parallel-transport condition. The imaginary part of the IGP of PTQM affects the thermal weights and introduces effective temperatures.
Consequently, even in a simple two-level system, the IGP of PTQM can display interesting behaviors unavailable in conventional QM, such as the geometric phase transition of a two-level system at finite temperature. For more complicated non-Hermitian quantum systems, the generalized IGP may serve as a probe to uncover intriguing characteristics due to geometry and topology.

\section{Acknowledgments}
H.G. was supported by the National Natural Science
Foundation of China (Grant No. 12074064). X. Y. H. was supported by the Jiangsu Funding Program for Excellent Postdoctoral Talent (Grant No. 2023ZB611). C.C.C. was supported
by the National Science Foundation under Grant No.
PHY-2310656

\appendix
\section{Details of the geometric phase of pure states}\label{appa}
To derive the geometric phase shown in Eq.~(\ref{theta1t}), the expansion (\ref{Pt}) is plugged into Eq.~(\ref{eom2}), yielding
\begin{align}
\mi|\dot{\Psi}\rangle=\sum_m\left[(\mi \dot{c}_m+c_m\tilde{E}_m)\me^{\mi\theta^1_{\text{D}m}}|\Psi_m\rangle+c_m\me^{\mi\theta^1_{\text{D}m}}|\dot{\Psi}_m\rangle\right],
\end{align}
where $\tilde{E}_m=E_m+\mi\langle\Psi^0_m|S^{-1}\dot{S}|\Psi^0_m\rangle$. Applying Eq.~(\ref{eom2}), the left-hand-side becomes
\begin{align}
\mi|\dot{\Psi}\rangle=\sum_m c_m\me^{\mi\theta^1_{\text{D}m}}(E_m-\mi S\dot{S}^{-1})|\Psi_m\rangle.
\end{align}
Multiplying the above equations by $\langle\Phi_n|$ from the left and applying the relation \cite{JPAGW13}
\begin{align}
\langle \Phi_n|\dot{\Psi}_m\rangle=\frac{\langle \Phi_n|\dot{H}|\Psi_m\rangle}{E_n-E_m}\text{ for } m\neq m,
\end{align}
we get
\begin{align}
&\dot{c}_n=\mi c_n\langle\Phi_n|\dot{\Psi}_n\rangle\notag\\
+&\mi\sum_{m\neq n}c_m\me^{\mi(\theta^1_{\text{D}m}-\theta^1_{\text{D}n})}\langle\Phi_n|\left(\frac{\dot{H}}{E_n-E_m}+\mi S\dot{S}^{-1}\right)|\Psi_m\rangle.
\end{align}
As in the conventional quantum mechanics, the adiabatic approximation is employed, so level-crossing terms (i.e., terms with $m\neq n$) are dropped. We finally get
\begin{align}
\dot{c}_n(t)\doteq\mi c_n(t)\langle\Phi_n|\frac{\dif}{\dif t}|\Psi_n\rangle,
\end{align}
whose solution is
\begin{align}
c_n(t)\approx c_n(0)\me^{-\int_0^t\dif t'\langle \Phi_n(t')|\frac{\dif}{\dif t'}|\Psi_n(t')\rangle}.
\end{align}

Next, we verify that
\begin{align}\label{thetaBn3}
\theta^2_n(\tau)=&\mi\oint_C\dif t\langle \Psi^0_n(t)|\frac{\dif}{\dif t}|\Psi^0_n(t)\rangle\notag\\=&\mi\oint_C\dif \mathbf{R}\cdot\left[\left\langle\Psi_n|W \nabla| \Psi_n\right\rangle+\frac{1}{2}\left\langle\Psi_n|(\nabla W)| \Psi_n\right\rangle\right]\notag\\
=&\theta^\text{B}_n
\end{align}
subject to $\dot{S}^{-1}S=(\dot{S}^{-1}S)^\dag$.
Using $|\Phi_n(t)\rangle=W(t)|\Psi_n(t)\rangle$ and $W^\dag=W$, the first term on the right-hand-side of Eq.~(\ref{thetaBn1}) is nothing but $\theta^1_n$, which can be further expressed as
\begin{align} \label{Berry}
&\theta^1_n =\mi \oint \dif t\langle\Phi_n(t)|\frac{\dif}{\dif t}| \Psi_n(t)\rangle  \notag\\
=&\mi \oint\dif t\langle\Psi^0_n(t)|S^{-1} \dot{S}| \Psi^0_n(t)\rangle +\mi \oint \dif t\langle\Psi^0_n(t)|\frac{\dif}{\dif t}| \Psi^0_n(t)\rangle\notag\\
=&\mi \oint\dif t\langle\Psi^0_n(t)|S^{-1} \dot{S}| \Psi^0_n(t)\rangle +\theta^2_n.
\end{align}
The second term on the right-hand-side of Eq.~(\ref{thetaBn1}) is
\begin{align}\label{geometry}
&\frac{\mi}{2}\oint_C\langle\Psi_n(t)|\dot{W} | \Psi_n(t)\rangle \dif t,\notag\\
	=&\frac{\mi}{2}\oint_C\langle\Psi_n(t)|(\dot{S}^{-1})^\dagger S^{-1} + (S^{-1})^\dagger \dot{S}^{-1}| \Psi_n(t)\rangle \dif t,\notag\\
	=&\frac{\mi}{2}\oint_C\langle\Psi^0_n(t)|\left[S^\dagger(\dot{S}^{-1})^\dagger  + \dot{S}^{-1}S\right]| \Psi^0_n(t)\rangle \dif t,\notag\\
	=&-\mi \oint_C\langle\Psi^0_n(t)| S^{-1}\dot{S}| \Psi^0_n(t)\rangle \dif t,
\end{align}
where we have applied $S^\dagger(\dot{S}^{-1})^\dagger = \dot{S}^{-1}S$ from the proper mapping condition. Along with Eq.~(\ref{Berry}), we conclude that $ \theta^\text{B}_n= \theta^\text{2}_n$.

\section{Details of geometric phase of thermal states}\label{appb}
To verify that $U(t)$ in Eq.~(\ref{Ut}) satisfies the parallel-transport condition (\ref{pxcmnh1}), we need the following identities:
\begin{align}
\dot{U}(t)  =&-\sum_n\langle\Phi_n(t)|\frac{\dif}{\dif t}| \Psi_n(t)\rangle U(t)\notag\\+&\sum_n \me^{- \int \langle\Phi_n(t')|\frac{\dif}{\dif t'}| \Psi_n(t')\rangle \dif t'} \left(\frac{\dif}{\dif t}|\Psi_n(t)\rangle\right) \langle\Phi_n(0)|,\notag\\
U^{-1}(t)&=\sum_{n} \me^{\int  \langle\Phi_n(t') |\frac{\dif}{\dif t'} | \Psi_n(t') \rangle\dif t'}\left|\Psi_n(0)\right\rangle\left\langle\Phi_n(t)\right|.\notag
\end{align}
They lead to Eq.~(\ref{pxcmnh1}):
\begin{widetext}
\begin{align}
\langle \Phi_n|\dot{U}(t)U^{-1}(t)|\Psi_n\rangle
		& =\langle \Phi_n(t)| \Psi_n(t)\rangle \langle\Phi_n(t)|    \bigg[ -\langle\Phi_n(t)|\frac{\dif}{\dif t}| \Psi_n(t)\rangle  |\Psi_n(t) \rangle \langle\Phi_n(t) |+  \left(\frac{\dif}{\dif t} |\Psi_n(t) \rangle\right)  \langle\Phi_n(t) | \,  \bigg]  | \Psi_n(t) \rangle \notag\\
	& = -\langle\Phi_n(t)|\frac{\dif}{\dif t}| \Psi_n(t)\rangle + \langle\Phi_n(t)|\frac{\dif}{\dif t}| \Psi_n(t)\rangle \notag\\
	& =0.
\end{align}
\end{widetext}

\section{Proper mapping of the two-level system}\label{appc}
To search for a proper mapping $S$ of our example in the main text, we first notice that $W=(S^{-1})^\dag S^{-1}$, which is invariant under a $U(N)$ transformation $u$:  $S^{\prime-1}=uS^{-1}\rightarrow W=(S^{\prime-1})^{\dagger} S^{\prime-1}$. We can use this degree of freedom to obtain a proper $S$.
For convenience, we initially take $W=(S^{-1})^2$ or conversely $S^{-1}=\sqrt{W}$ since $W$ is already given by Eq.~(\ref{W}). Since $W$ is Hermitian, this kind of $S$ has at least two solutions:
\begin{align}\label{similarity}
	S^{-1}_{\pm}=
	\frac{\begin{pmatrix}
		\sqrt{a^2-b^2}\pm a & \mi b \me^{-\mi \pi} \\
		-\mi b \me^{\mi \pi} & \sqrt{a^2-b^2}\pm a \\
	\end{pmatrix}}{\sqrt{2 a \left(\sqrt{a^2-b^2}\pm a\right)}}	,
\end{align}
Take $a=3$ and $b=\sqrt{5}$ and choose $S=S_{+}$ without loss of generality, then
\begin{align}
	S=\begin{pmatrix}
		\frac{\sqrt{\frac{15}{2}}}{2} & -\frac{1}{2} \mi \sqrt{\frac{3}{2}} \me^{-\mi \phi } \\
		\frac{1}{2} \mi \sqrt{\frac{3}{2}} \me^{\mi \phi } & \frac{\sqrt{\frac{15}{2}}}{2}
	\end{pmatrix}
	\end{align}
and the original Hamiltonian is converted to
\begin{align}
H_0&=S^{-1} H S=\begin{pmatrix}
		\epsilon +2 \cos \theta  & 2 \me^{-\mi \phi } \sin \theta  \\
		2 \me^{\mi \phi } \sin \theta  & \epsilon -2 \cos \theta
	\end{pmatrix}.
	\end{align}
The eigenvector of $H_0$ associated with $E_+=\epsilon+2$ is
\begin{align}
	| \Psi^0_+(\theta,\phi)\rangle=\begin{pmatrix}
		\frac{\me^{-\mi \phi } (\cot \theta +\csc \theta )}{\sqrt{(\cot \theta +\csc \theta )^2+1}}\\
		\frac{1}{\sqrt{(\cot \theta +\csc \theta )^2+1}}
	\end{pmatrix},
\end{align}
which leads to
\begin{align}\label{inte}
	&\langle \Psi^0_+| 	S^{-1} \dif S | \Psi^0_+\rangle =-\frac{1}{4}  (\sqrt{5} \sin \theta +\mi \cos \theta )\dif\phi,\notag\\
	&\mi \oint_C\langle \Psi^0_+| 	S^{-1} \dif S | \Psi^0_+\rangle \dif t=\frac{\pi}{2}(\cos \theta -\mi\sqrt{5} \sin \theta ).
\end{align}
Apparently, the second term of $\theta^1_+$ is complex in this case. To make it purely imaginary, we impose a unitary transformation $S^{-1}= u S^{-1}_{\text {proper }}$, where $u$ can be fixed by the condition of a proper mapping $\dot{S}^{-1}_{\text {proper }} S_{\text {proper }}=(\dot{S}^{-1}_{\text {proper }} S_{\text {proper }})^{\dagger}$. This is equivalent to solving the equation
\begin{align}
	\dot{u}=\frac{1}{2}\left[\dot{S}^{-1} S-\left(\dot{S}^{-1} S\right)^{\dagger}\right] u
\end{align}
subject to the initial condition $u(0)=\mathbf{1}_{2\times 2}$.
The general solution is quite involved. Fortunately, if the system evolves along a circle of latitude such that $\dif \theta=0$, an analytical expression of $u$ can be found as
\begin{align}
	u(\phi)=\begin{pmatrix}
		\me^{\frac{\mi  \phi }{4}} & 0 \\
		0 & \me^{-\frac{\mi \phi }{4}}
	\end{pmatrix}.
\end{align}
Accordingly, the proper mapping $S$ is
\begin{align}
	S_{\text {proper }}=Su=\begin{pmatrix}
		\frac{1}{2} \sqrt{\frac{15}{2}} \me^{\frac{\mi \phi }{4}} & -\frac{1}{2} \mi \sqrt{\frac{3}{2}} \me^{-\frac{5 \mi \phi}{4} } \\
		\frac{1}{2} \mi \sqrt{\frac{3}{2}} \me^{\frac{5 \mi \phi }{4}} & \frac{1}{2} \sqrt{\frac{15}{2}} \me^{-\frac{\mi \phi}{4} } \\
	\end{pmatrix}.
\end{align}

\bibliographystyle{apsrev}
\bibliography{Review1}

\end{document}